\documentclass[aps,prl,reprint,groupedaddress]{revtex4-1}

\usepackage{txfonts}
\usepackage{amssymb}
\usepackage{bm}
\usepackage[pdftex]{graphicx}
\usepackage{color}
\usepackage{longtable}

\usepackage{url}
\usepackage[colorlinks=true, linkcolor=blue, citecolor=blue, filecolor=blue, urlcolor=blue]{hyperref}
\usepackage{siunitx}

\begin{document}


\title{Chirality-Induced Selectivity of Phonon Angular Momenta in Chiral Quartz Crystals}


\author{Kazuki Ohe$^{1}$}
\author{Hiroaki Shishido$^{1,2}$}
\author{Masaki Kato$^{3}$}
\author{Shoyo Utsumi$^{2}$}
\author{Hiroyasu Matsuura$^{3}$}
\author{Yoshihiko Togawa$^{1,2,4}$}
\email[]{ytogawa@omu.ac.jp}
\affiliation{$^1$Department of Physics and Electronics, Osaka Prefecture University, 1-1 Gakuencho, Sakai, Osaka 599-8531, Japan}
\affiliation{$^2$Department of Physics and Electronics, Osaka Metroplitan University, 1-1 Gakuencho, Sakai, Osaka 599-8531, Japan}
\affiliation{$^3$Department of Physics, The University of Tokyo, Bunkyo, Tokyo 113-0033, Japan}
\affiliation{$^4$Quantum Research Center for Chirality, Institute for Molecular Science, Okazaki 444-8585, Japan}

\date{Received 8 April 2022; Revised 5 November 2022 and 18 October 2023.}

\begin{abstract}
A generation, propagation, and transfer of phonon angular momenta are examined on thermal transport in chiral insulative and diamagnetic crystals of $\alpha$-quartz. 
We found that thermally-driven phonons carry chirality-dependent angular momenta in the quartz crystals and they could be extracted from the quartz as a spin signal.
Namely, chirality-induced selectivity of phonon angular momenta is realized in the chiral quartz. 
We argue that chiral phonons available in chiral materials could be a key element in triggering or enhancing chirality-induced spin selectivity with robust spin polarization and long-range spin transport found in various chiral materials.
\end{abstract}


\maketitle


\textcolor{black}{Dynamical aspects of chiral objects have attracted much attention in a wide range of research fields~\cite{Barron_text, Wagniere2007}. As exemplified by studies in biochemistry~\cite{Hendry_2010}, nano-optics~\cite{Kelly_2020}, phonon~\cite{Kishine_2020}, and magnetism~\cite{Togawa_2012}, it is evident that a rotational motion accompanies a translational motion and vice versa in chiral objects. These features are well captured by the concept of dynamical chirality. An interplay between structural and dynamical chirality plays a key role in various chirality-induced phenomena~\cite{Togawa_Review2016, Togawa_Review2023}. 
In this respect, chirality-induced spin selectivity (CISS)~\cite{Gohler_2011} has been a matter of interest since the CISS satisfies the symmetry of dynamical chirality.}

\textcolor{black}{The CISS phenomena were initially found in non-conductive chiral molecules, in which electrons exhibit spin polarization in a nonlinear regime of tunneling transport~\cite{Gohler_2011, Xie_2011, Suda_2019}. Interestingly, the CISS signal increases as temperature increases around room temperature~\cite{Mondal_ACSnano_2020, Das_JPCC_2022}, stimulating discussions on a possible scenario of phonon-mediated spin polarization in chiral molecules~\cite{Du_2020, Zhang_2020, Fransson_2020, Kato_2022}. However, no direct evidence of phonons' contribution to the CISS has been obtained in the experiments yet.}

\textcolor{black}{Recently, it is found that chiral metallic crystals show a spin-polarized state in a linear regime of transport of conduction electrons~\cite{Inui_2020, Shiota_2021, Shishido_2021, Shishido_2023}. Importantly, a robust protection of the spin polarization occurs in chiral metallic crystals and a nonlocal electrical transport preserves the spin polarization over micrometers and even millimeters. These studies also lead to interesting arguments on a role of conduction electrons and phonons underlying the CISS~\cite{Kishine_2020}.}

\textcolor{black}{A quartz is an electrically insulative (precisely dielectric) and non-magnetic (precisely diamagnetic) material. Thus, there are no conduction and localized electrons that contribute to spin polarized phenomena. Indeed, a spinless electronic structure of chiral quartz is drastically different from those of chiral molecules and of chiral metallic crystals, providing good contrast with the previous studies. Furthermore, phonon properties of chiral quartz are well defined over the crystal~\cite{Pine_1969}, while magnons are absent, and would be in favor of realizing a long-range spin transport, as observed in chiral metallic crystals~\cite{Inui_2020, Shiota_2021, Shishido_2021, Shishido_2023}. In this respect, detecting the selectivity of chirality-dependent angular momenta in quartz is worth studying and may clarify a role of phonons in the CISS.}

\textcolor{black}{\textcolor{black}{In this Letter,} we study thermal transport in chiral crystals of $\alpha$-quartz in order to examine a generation, propagation, and transfer of angular momenta. We found experimentally that the phonons carry chirality-dependent angular momenta in the chiral quartz crystals and they could be extracted from the quartz as a spin signal. Analytical calculations of thermal transport with phonons revealed a thermally-driven generation and propagation of the net angular momenta in the chiral quartz via specific phonons, referred to as chiral phonons~\cite{Kishine_2020}.}

\begin{figure*}[t]
\includegraphics[width=1.0\linewidth]{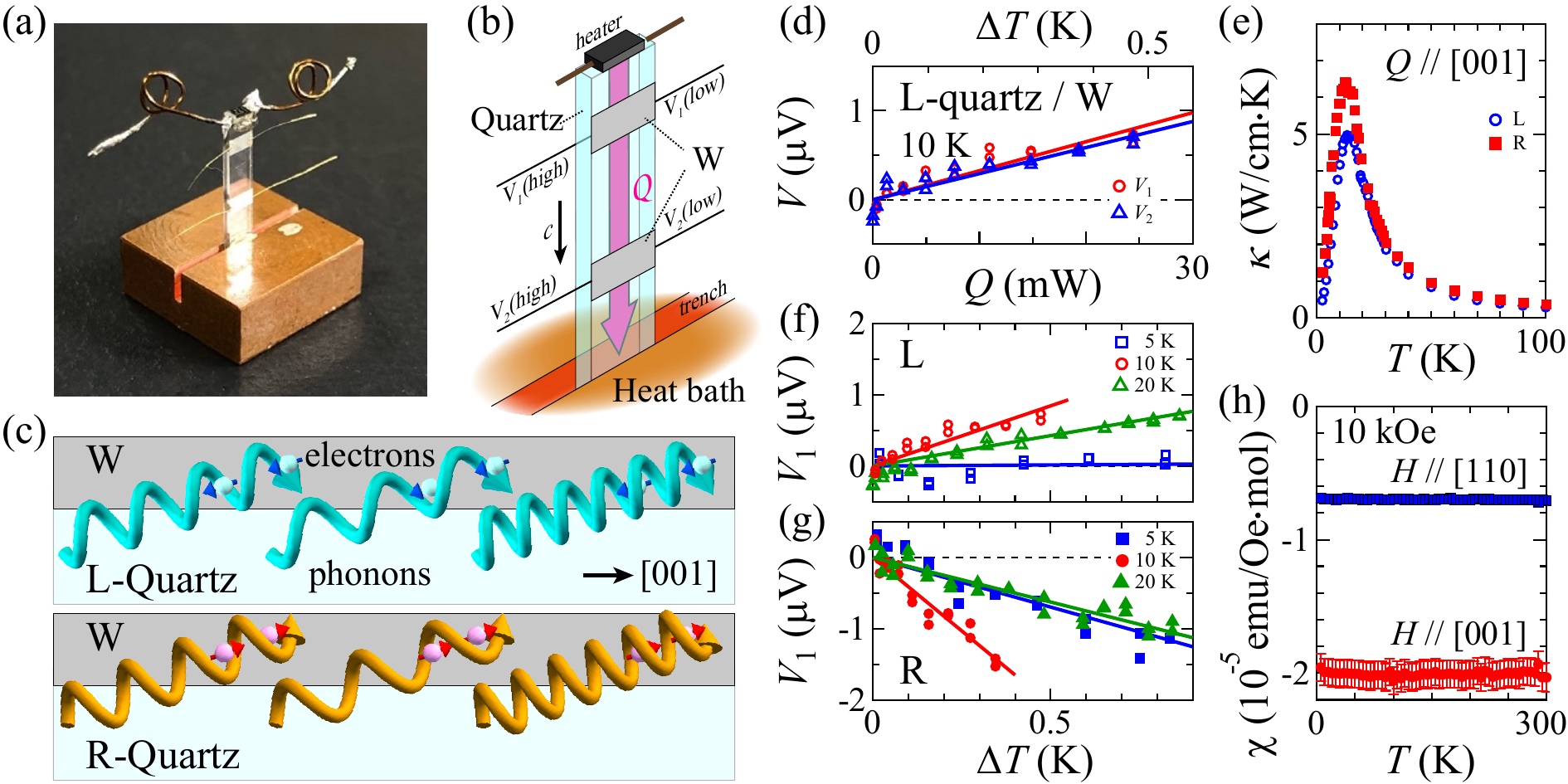}
\centering
\caption{
A photograph (a) and schematic (b) of the device, made of a quartz substrate with tungsten (W) electrodes. The size of a copper heat bath is 10 $\times$ 10 $\times$ \SI{4}{\milli \metre^{3}} with a trench of \SI{2}{\milli \metre} in depth and \SI{0.5}{\milli \metre} in width. 
The transverse voltage is detected along the electrodes located at the upper and lower sides of the substrate. 
(c)~A schematic for the propagating and leaking phonons that interact with electrons in the electrode on the quartz. 
(d)~The voltage as functions of $Q$ and $\Delta T$ at the heat-bath temperature of \SI{10}{\kelvin}. $\Delta T$ is estimated using $\kappa$ in (e), measured with the quartz substrates used in this study. (f) and (g)~The voltage at 5, 10, and \SI{20}{\kelvin} in the L/R-quartz. 
(h)~Diamagnetic magnetization of quartz.
}
\label{fig1}
\end{figure*}

\textcolor{black}{Quartz is a crystal of silicon dioxide SiO$_{2}$. Trigonal crystals of quartz, which are called $\alpha$-quartz, belong to the space group of $P$3$_{1}$21 or $P$3$_{2}$21. The former crystal (called R-quartz in this study) has a right-handed helical structure of SiO$_{4}$ tetrahedrons, while the latter (L-quartz) exhibits a left-handed helical one~\cite{quartz}. 
Single crystals of R/L-quartz were used in the present experiments. The size of the quartz substrates is \SI{2}{\milli \metre} in width ($w$), \SI{10}{\milli \metre} in length ($l$), and \SI{0.5}{\milli \metre} in depth ($t$). In most of the experiments, the helical $c$-axis of quartz is oriented in the direction along the longest dimension of the substrate, while the heat flow direction is fixed along that direction.}

\textcolor{black}{In the device designed for thermal transport measurements, which follows the experimental setup used for the conventional thermal measurements at cryogenic temperatures (Fig.~S1)~\cite{White_textbook_1979}, the quartz substrate stands up on a heat bath made of copper with a trench of \SI{2}{\milli \metre} in depth ($d$) for thermal contact, while a tip resistor is put on the top surface of the substrate as an electrical heater, as shown in Fig.~\ref{fig1}(a). 
An excitation of input thermal power $Q$ with a small magnitude is set as a source of temperature difference $\Delta T$.
This measurement setup guarantees an introduction of temperature gradient $\nabla T$ from the top surface to the bottom one of the substrate.}

\textcolor{black}{For the signal detection, two electrodes of tungsten (W) or platinum (Pt), which has large spin-orbit coupling and spin Hall angles with the opposite sign~\cite{wang2014, niimi2015}, were fabricated with a separation length of \textcolor{black}{\SI{2.8}{\milli \metre}} by photolithography and sputtering methods. A thickness of the electrodes is optimized to be \SI{6}{\nano \metre} for both W and Pt so as to enlarge the signal generated by spin-charge conversion in the electrode. The signal is also examined in another device with the single W electrode of different thickness, \textcolor{black}{which is located at the center of the quartz substrate along the $l$ direction}.}

\textcolor{black}{Let us conceive how the device works. When the electrical heater turns on, the thermal current is induced through the quartz substrate. We assume that electrons in the electrode locating on the way to the thermal bath would absorb angular momenta, carried by the propagating phonons that leak into the electrode from the substrate, via a phonon-drag process~\cite{Wang2013, Kimura2021}. 
Another assumption made here is that such a diffusive flow of angular momenta into the electrode could be regarded as a spin diffusive current via spin-phonon coupling. Then, it is converted into a transverse charge current via an inverse spin Hall effect (ISHE)~\cite{valenzuela2006, saitoh2006, kimura2007} and finally detected as a voltage ($V$) output along the electrode~\cite{Inui_2020, Shiota_2021, Shishido_2021, Shishido_2023}.}

\textcolor{black}{Figure~\ref{fig1}(d) shows the $V$ signals generated in the W detection electrodes at the upper and lower sides of the quartz with the heat-bath temperature fixed at \SI{10}{\kelvin}. They increase linearly with almost the same magnitude against $Q$ and $\Delta T$.
Here, $\Delta T$ was estimated using the thermal conductivity $\kappa$ measured with the quartz used in the present study [Fig.~\ref{fig1}(e)].
This result indicates that the thermal current is induced uniformly at least over \SI{3}{\milli \metre} in the quartz substrate beneath the electrodes. 
In addition, propagating and leaking phonons interact with electrons prevailing in the electrodes and generate the spin signal as the $V$ output.}

\begin{figure*}[t]
\includegraphics[width=1.0\linewidth]{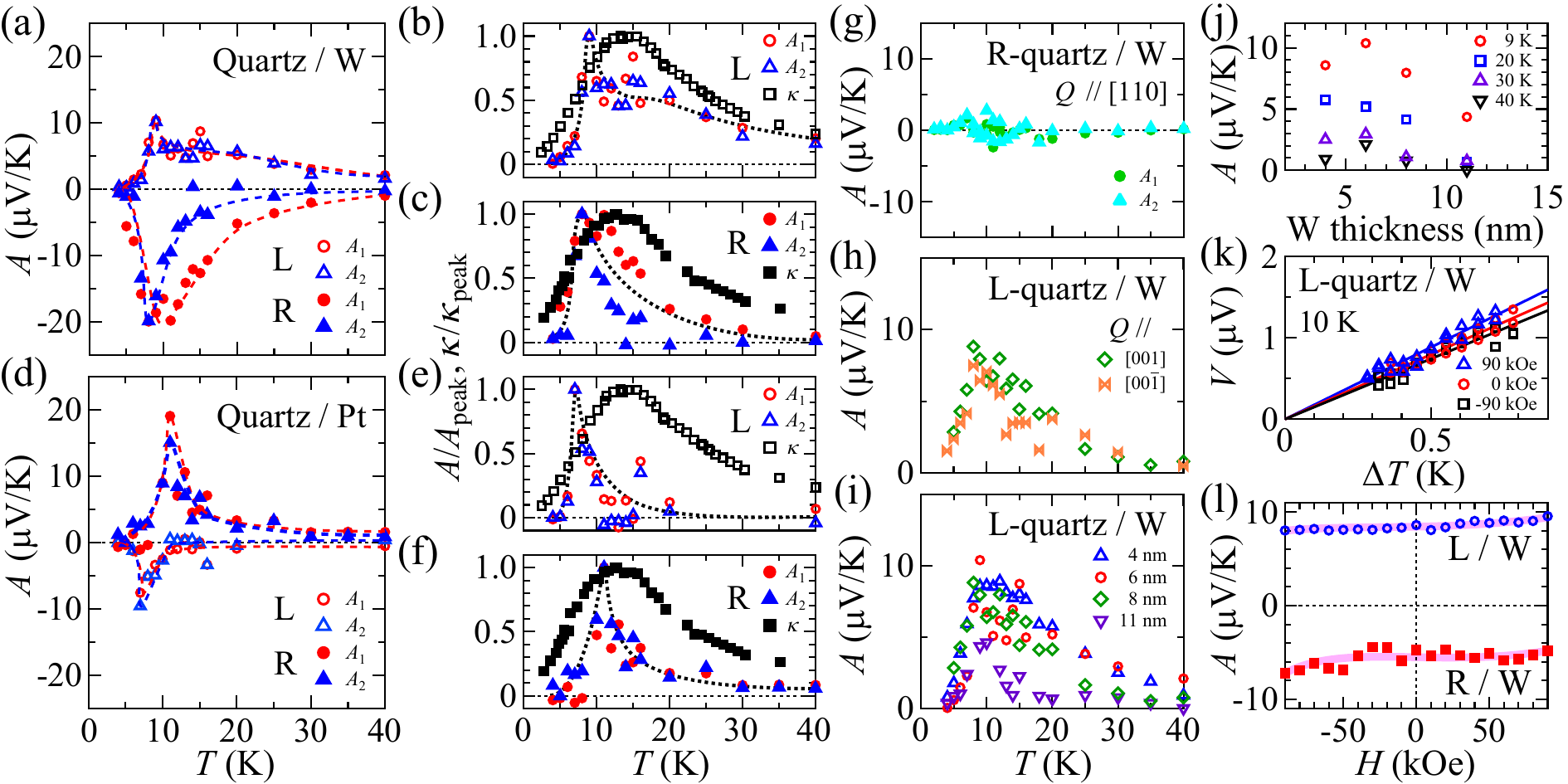}
\centering
\caption{
(a) \textcolor{black}{$T$} dependence of conversion coefficient $A$ in the L/R-quartz with W electrodes.
(b) and (c)~Normalized $A$ with the reference of $\kappa$ measured for the quartz.
(d) to (f) The dataset for the L/R-quartz with Pt. 
(g) Anisotropic signal under the heat flow along [110] in the R-quartz with W.
(h) Control experiment with the heat flow reversed from [001] to [00$\overline{1}$]. 
(i) and~(j) Dependence of $A$ on the W thickness. 
(k) and~(l) The response under $H$ along the $c$-axis at \SI{10}{\kelvin}.
\textcolor{black}{The data in (h)-(l) is taken at the center of quartz except for the $A_{1}$ data for \SI{6}{\nano \metre} in (i) and (j).}}
\label{fig2}
\end{figure*}

\textcolor{black}{To confirm this interpretation, the $V$ signals were examined at various heat-bath temperatures in the quartz with the opposite handedness. Figures~\ref{fig1}(f) and~\ref{fig1}(g) show such data for the L/R-quartz substrates, respectively. It is clear that the signal reverses its sign depending on the handedness of quartz. Furthermore, it is found that the slope of the signal is steepest around \SI{10}{\kelvin} with regard to thermal-bath temperatures.}

\textcolor{black}{A conversion coefficient $A$ is given by the ratio of the electric field $E$ and $\nabla T$. Namely, $A \equiv E/\nabla T = (V/\Delta T)(l-d)/w$, which follows the definition of the spin Seebeck coefficient~\cite{Uchida_2008, Jaworski_2008}. $A$ is shown as a function of thermal bath temperatures in Fig.~\ref{fig2}(a). Apparently, the peak structure is observed at approximately \SI{10}{\kelvin} in all the data, while the coefficient sign alters reflecting the handedness of quartz. The coefficient magnitude reaches the order of tens of microvolts per Kelvin in the quartz, which is as large as typical values of the spin Seebeck coefficient for magnetic materials~\cite{Uchida_2008}.}

\textcolor{black}{When normalizing the data by the peak value, the universal behavior is clearly seen in Figs.~\ref{fig2}(b) and~\ref{fig2}(c). Note that $\kappa$ takes the maximum at around \SI{10}{\kelvin} in the quartz crystals, as observed in the literature~\cite{Zeller_1971}. Interestingly, the experimental data follows this tendency of the quartz.}

\textcolor{black}{Regarding quasi particles responsible for thermal transport in the quartz, only the phonons are available since neither conduction electrons nor magnons exist due to insulative and diamagnetic nature of quartz. In such a case, $\kappa$ is given by the following equation $\kappa = \frac{1}{3} C v l$, where $C$ is a specific heat, $v$ is a drift velocity of phonons, and $l$ is a mean free path of phonons. With decreasing temperature, $\kappa$ becomes larger since $l$ extends. However, at much lower temperatures, a population of phonons declines and thus $\kappa$ becomes small through a reduction of $C$. Such a peak structure is clearly observed in the quartz crystals, showing the maximum value of $\kappa$ around \SI{10}{\kelvin}. Thus, the coincidence between the data and $\kappa$ behavior strongly supports that the spin voltage is originated from the phenomena governed by phonon propagation and angular momenta transfer.}

\textcolor{black}{In this connection, it is worth mentioning that thermoelectric materials exhibit the coincidence of a temperature dependence of thermoelectric Seebeck coefficient and $\kappa$~\cite{Wang2013, Kimura2021, Bentien_2007,Takahashi_2016}. This coincidence is regarded as a manifestation of the phonon-drag effect, in which the flow of charge carriers is dragged by the heat flow of phonons via phonon-electron interaction~\cite{Gurevich_1945, Ziman_text, Battiato_2015, Matsuura_2019}. Even in the present experiments, such an interplay between leaking phonons and electrons may be involved through the interface between the insulative quartz and metallic electrodes, as indicated by the data in Fig.~\ref{fig2}.} 

\textcolor{black}{To see the nature of the signals generated in the detection electrodes, the electrode material is replaced from W to Pt. Then, it is found that the relationship reverses completely with regard to the electrode elements and handedness of quartz, as shown in Figs.~\ref{fig2}(d) to~\ref{fig2}(f). The difference in the peak height and position may reflect the quality of L/R-quartz crystals.}

\textcolor{black}{If the $V$ signal observed at \SI{10}{\kelvin} arises from the conventional Seebeck effect in the metallic electrodes~\cite{Moore_1973}, $\Delta T_{trans}$ of more than \SI{1}{\kelvin} is required in the transverse direction across the quartz substrate, which was hardly induced in the present experiment with an excitation of $\Delta T$ less than \SI{0.6}{\kelvin} along the longitudinal direction taking into account a good value of $\kappa$ for quartz and negligibly small thermal radiation at \SI{10}{\kelvin}. In this connection, a longitudinal spin Seebeck effect~\cite{Meier_2015}, influenced by $\Delta T$ normal to the substrate toward the electrode, hardly occurred in the present experiments either.}

\textcolor{black}{An application of external magnetic field $H$ has no influence on the insulative quartz, as evidenced by its diamagnetic behavior [Fig.~\ref{fig1}(h)]. This feature gives strong contrast with spin transport experiments with magnetic materials, in which the control of magnetic moments' orientation using $H$ is inevitable to obtain finite signals. In the present setup, the presence of $H$ perpendicular to the heat flow induces a Nernst effect in the metallic electrodes. Thus, to avoid artefacts, most of the experiments were performed without $H$.}

\textcolor{black}{We stress that a lack of conduction electrons and magnons in insulative and diamagnetic quartz limits possible mechanisms responsible for spin transport in the quartz. For instance, anomalous Nernst and spin Nernst effects~\cite{Meyer_2017, Bosea_2019} are hardly involved in the phenomena occurring in the quartz.}

\textcolor{black}{A control experiment was carried out in the same measurement setup with a magnesium oxide substrate (Fig.~S2). No emergence of the signal in the achiral (non-chiral) substrate with the detection electrodes guarantees that the observed phenomena are relevant to the chirality of quartz.}

\textcolor{black}{Thermal transport was also examined in the quartz with the $c$-axis orienting in the transverse direction of the platelet substrate [Fig.~\ref{fig2}(g)]. The spin signal reduces significantly when $\Delta T$ is applied in the [110] direction.}

\textcolor{black}{The signal sign should be opposite when reversing $\Delta T$ because of the two-fold symmetry within the $c$-plane of quartz. In the experiment, the signal was examined with the configuration of quartz reversed in the same device setup of the single W electrode and of the direction of heat flow. \textcolor{black}{This setup gives the data collected at the substrate center.}  
Almost the same data was obtained for the opposite heat flow [Fig.~\ref{fig2}(h)], as consistent with the symmetry of quartz. \textcolor{black}{Here, we stress that the conversion coefficients exhibit the same sign and almost the same magnitude at three different locations [Figs.~\ref{fig2}(a) and~\ref{fig2}(h)].}}

\textcolor{black}{The data was also taken with the W electrode of different thickness [Figs.~\ref{fig2}(i) and~\ref{fig2}(j)].  A monotonic increase appears with reducing the thickness, followed by a peak structure at small thickness. This data is in agreement with typical behavior of the ISHE~\cite{Castel_2012, Hahn_2013} and rules out the contribution of the Seebeck signal, which is determined by $\Delta T$ and independent of electrode thickness.} 

\textcolor{black}{Signal reduction and enhancement were found when applying an in-plane $H$ of \SI{\pm9}{\tesla} along the $c$-axis, while the signals were less dependent at small $H$ [Figs.~\ref{fig2}(k) and~\ref{fig2}(l)]. Such a relationship reverses in terms of the chirality. This behavior indicates a robust generation of spins, which tilt with regard to the $c$-axis, as schematically drawn in Fig.~\ref{fig1}(c), and thus exhibit the precession dynamics at large $H$.}

\textcolor{black}{Using the same devices (without the tip resistor), we have tried to make a spin injection from the transverse electrode into the chiral quartz via a spin Hall effect~\cite{kimura2007, Inui_2020}. In the experiment, while the current was applied to the electrode on the upper side of the quartz, the $V$ signal was detected on the other electrode on the lower side of quartz. An odd component did not appear in current-voltage characteristics, while an even component was observed (Fig.~S3), suggesting that the signal is generated by a thermally-driven conversion process in the chiral quartz. Indeed, even in this experimental setup, the signal sign reverses in the quartz with the opposite handedness, which is consistent with the data in Figs.~\ref{fig1} and~\ref{fig2}.}  

\textcolor{black}{
The output voltage signal is definitely stemmed from a spin conversion process via the ISHE in the detection electrodes. 
It turns out that, based on the experimental observations, W and Pt elements with the spin Hall angle of the opposite sign are sensitive to a sign of angular momenta flowing from the chiral quartz into the electrode. 
Note that the signal sign is dependent on the chirality of quartz. 
Therefore, we can conclude that angular momenta with the opposite sign are generated in the quartz with the different handedness, predominantly propagate through it, and are transferred from it into the electrode. 
Namely, the present study demonstrates successfully chirality-induced selectivity of angular momenta along the phonon propagation in the chiral quartz crystals.}

\begin{figure}[t]
\includegraphics[width=1.0\linewidth]{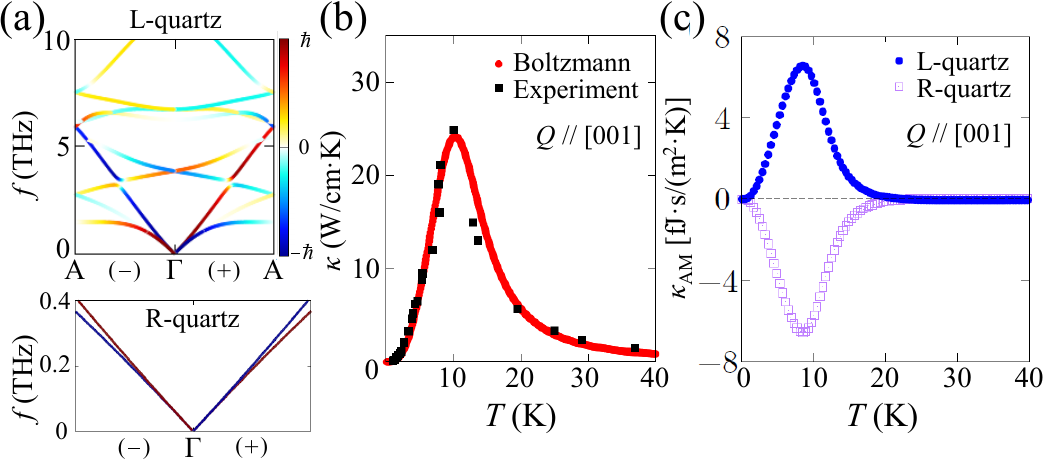}
\caption{Analytical calculations of phonon dispersion (a), $\kappa$ (b), and propagation of net angular momenta (c) for quartz. The $\kappa$ data is derived from the literature~\cite{Zeller_1971}.}
\label{fig3}
\end{figure}

\textcolor{black}{To validate the scenario experimentally obtained, the generation and propagation of phonon angular momenta were numerically examined in terms of the phonon dispersion of $\alpha$-quartz. A split of the phonon branches appears in the vicinity of $\Gamma$ point along the $\Gamma$-A line. Importantly, two of the phonon branches carry the angular momenta with the opposite sign. The sign depends on the handedness of the quartz. 
Then, thermal transport was analyzed on the basis of the Boltzmann equation of phonons with a relaxation time approximation, as explained in the supplemental material.
Figure~\ref{fig3}(c) shows the net angular momenta carried by thermally-driven phonons at a finite temperature. A peak structure appears at around \SI{10}{\kelvin} in a temperature dependence of the coefficient for the angular momenta propagation. This result is quite consistent with the experimental data in Fig.~\ref{fig2}.}

\textcolor{black}{Quite recently, the presence of chiral phonons unique to chiral materials~\cite{Kishine_2020} has been demonstrated experimentally in a chiral insulator of $\alpha$-HgS~\cite{Ishito_2022} and other materials~\cite{Ishito_2022b, Tsunetsugu_2023, Oishi_2022, Ueda_2023}.
The chiral phonons have two enantiomeric modes that are propagating with a finite group velocity while atoms are ellipsoidally rotating. A nonreciprocal propagation of the chiral phonons occurs with the generation of net angular momenta under $\nabla T$ because of a difference of the group velocity of enantiomeric chiral phonons, as shown in Fig.~\ref{fig3}. These features of the chiral phonons can explain the experimental results of chirality-induced selectivity of angular momenta along the phonon propagation in the chiral quartz. The thermally-driven process of phonon angular momenta is also discussed in the literature~\cite{Chen_2022, Hamada_2018}.}

\textcolor{black}{The data was successfully observed at temperatures up to around \SI{50}{\kelvin} in the present experiments. The data at room temperature sounds intriguing but was not observed convincingly since the enhancement of thermal radiation in a range of such high temperatures made it hard to get reliable data and identify the main contributor in thermal transport. Here we do point out a robustness of the electronic band structure of quartz against thermal agitation. Quartz crystals have an energy gap of \SI{8.4}{e\volt}~\cite{Philipp_1985}, which is approximately a thousand times larger than a thermal activation energy of the temperatures for the experiments. Thus, thermally activated electrons do not join diffusive thermal transport even at room temperature. An acquisition of convincing data at room temperature is left as future work.} 

\textcolor{black}{In the present study, we have investigated spin polarization phenomena in the thermal transport using chiral insulative quartz. It would be interesting to perform similar experiments in other chiral materials in terms of a correlation with the electrical conductivity of chiral materials. Phonons and electrons are in charge of thermal transport in chiral metallic materials, where the conventional CISS effect~\cite{Inui_2020, Shiota_2021, Shishido_2021, Shishido_2023} could emerge as well. Spin polarization generated in the electronic drift current via the CISS may amplify the generation of spin angular momenta of the chiral phonons. Taking advantage of a long coherence of the phonon propagation with a long wavelength, transferring the angular momenta from the chiral phonon to electron spins is likely to be realized on a macroscopic scale. Further experimental and theoretical considerations on phonon-mediated transport of spin angular momenta will clarify the mystery of the robust CISS response over macroscopic distances at room temperature found in different categories of chiral materials.}

\begin{acknowledgments}
We thank Takuya Satoh, Yusuke Kato, Yusuke Kousaka, Hiroshi M. Yamamoto, Masao Ogata, Yasuhiro Niimi, Hiroaki Kusunose, Jun-ichiro Ohe, and Jun-ichiro Kishine for useful discussions. We acknowledge support from Grants-in-Aid for Scientific Research (Grant Nos. 17H02767, 17H02923, 19K03751, 21H01032, 22H01944, and 23H00091) and Research Grant of Specially Promoted Research Program by Toyota RIKEN. A part of this work was conducted in Equipment Development Center (Institute for Molecular Science), supported by Nanotechnology Platform Program (Molecule and Material Synthesis) of the Ministry of Education, Culture, Sport, Science and Technology (MEXT), Japan. This research was also supported by the grant of OML Project by the National Institutes of Natural Sciences (NINS program No. OML012301).\end{acknowledgments}

\end{document}